# Forming 1D Periodic J-aggregates by Mechanical Bending of BNNTs: Evidence of Activated Molecular Diffusion

J.-B. Marceau[1], D.-M Ta[2] A. Aguilar[2], A. Loiseau[3], R. Martel[4], P. Bon[2], R. Voituriez[5], G. Recher[1], and E. Gaufrès[1*]

[1] Laboratoire Photonique Numérique et Nanosciences, Institut d'Optique, CNRS UMR5298, Université de Bordeaux, F-33400 Talence, France

[2] XLIM UMR CNRS 7252, Limoges, France

[3] Laboratoire d'Étude des Microstructures, ONERA-CNRS, UMR104, Université Paris-Saclay, BP 72, 92322 Châtillon Cedex, France

[4] Département de chimie et Institut Courtois, Université de Montréal, Montréal, Québec H3C 3J7, Canada

[5] Laboratoire Jean Perrin et Laboratoire de Physique Théorique de la Matière Condensée CNRS Sorbonne Université, 4 place Jussieu, 75005 Paris, France

*Correspondence: etienne.gaufres@cnrs.fr

**Abstract**

**Driving molecular assembly into micrometer-scale patterns is key for defining advanced materials of interest in various fields, including life sciences, photovoltaics, and quantum photonics. However, the driving process competes with other forces, such as Brownian motion, ripening phenomena, capillary forces, and non-specific adsorption. Here we report on a guided diffusion mechanism of luminescent dye molecules encapsulated inside boron nitride nanotubes (BNNTs). Correlative measurements between BNNT bending and molecular position along the BNNT axis reveal an efficient and long-range migration of dyes from curved to straight regions of the nanotube. This curvature activated diffusion forms clusters of bright J-aggregates in periodic patterns of well-defined spacing and length. A phenomenological model of guided molecular transport in bended BNNTs is used to describe this directed 1D diffusion inside BNNT. It is shown to accurately predict the position and morphologies of a J-aggregate as a function of nanotube length. Coupling topological stimuli to 1D molecular diffusion at the nanoscale is here presented as an interesting tool capable of reconfiguring various emissive patterns of functional molecules at the mesoscopic scale.**

**Main:**

Controlling aggregation of molecules at the nanoscale is an important tool to induce patterns in advanced materials for fields as diverse as bioimaging, electronics, photovoltaics, quantum photonics, to name just a few. For instance, guided assembly has been applied to create superstructures of few hundreds of molecular dyes assembled as bright emitters, which are promising materials for the second quantum revolution based on photon quantum entanglement and Boson sampling effects. [1–6] Driving molecular motion to form advanced photonic superstructures is however challenging. While specific chemical interactions can be used to form well-defined assemblies on the nanometer scale, the assembly process in the van der Waals regime often drives other long-range processes that lack the required selectivity to induce larger mesoscale patterns. Brownian diffusion, ripening processes, capillary, and drift effects generate, for instance, a plethora of undesirable movements at the scale of a single object, which alter both the morphology and the optical properties of superstructures. The formed superstructures are also prone to irreversible degradation mechanisms, such as oxidation and photobleaching, and often need additional protection from the environment. This is particularly acute for light emitting organic dye aggregates, whose optical properties are extremely sensitive, not only to the type of aggregation but also to their surrounding dielectric and molecular disorder in the first few nanometers.

Various approaches have been developed to handle large statistical quantities of molecules. Templates made of polymeric matrices[7–9] or micelles[10] have given many examples of good results. Centrifugation and dielectrophoresis effects[11,12], as well as other activated diffusion processes of asymmetrically shaped nanostructures, such as Janus molecules, have also been described theoretically and experimentally to successfully drive self-organized patterns of higher complexity.[13–16] Other strategies enabling quasi-deterministic handling or patterning of molecular assemblies are now being searched for gaining even high control. Optical tweezers, for example, make it possible to manipulate matter at the micron scale, from single molecules to small aggregates, but a proper matching the refractive index between the tweezers and the nanoobjects is required.[17–19]. Molecular transport and diffusion mechanisms observed in life science can also be used to harness molecular motion. For instance, the diffusion and/or displacement of molecular motors on microtubules or intercellular tunneling nanotubes can manage the displacement of matter on a scale of hundreds of micron.[20–23] Both of these strategies are limited the scope: they mostly work in a dilute regime and they are not scalable and are poorly adapted for manipulating dense molecular assemblies.

Mixed statistical/deterministic strategies have been investigated in which a reservoir of molecules placed in a liquid or gaseous environment self-assembles spontaneously into aggregates or advanced supramolecular structures.[24] The engineering of complex phases has been developed, for example, using micrometer-sized periodic patterns of molecules, as observed in spinodal phase transitions.[25,26] Tailored molecular adsorption on smart templates, such as nanoporous materials made using Zeolites and Metal Organic Framework (MOF)[27,28] and functionalized surfaces, including DNA origami[29], further expands these mixed strategies for nanoscale organization. In this framework, nanotubes have attracted considerable interest as templates to assemble and protect a wide variety of molecular aggregates encapsulated inside their cavities or adsorbed on their outer surface.[30–37] Of relevance to this work, we have recently demonstrated that rod-like organic dyes confined and aligned in boron nitride nanotubes (BNNT) emit bright, stable and polarized light.[38,39]

Here we show how 1D molecular diffusion of organic dyes inside BNNTs can be driven by mechanical bending forces. More precisely, mechanical deformations of BNNTs were induced at the nanometer scale to induce bending movements. These deformations activate 1D molecular diffusion and aggregation into periodic locations inside the nanotube host. The phenomenon produces stable periodic emitting 1D clusters of molecules of length between 0.6 and 2 μm depending on the nanotube length. The underlying mechanism is studied and ascribed to local changes of the longitudinal curvature of BNNTs, which activate molecular diffusion in regions in the BNNTs of maximal curvature to the nearby sections of low curvature. A theoretical model based on the bending properties of BNNT combined to activated diffusion induced by curvature is proposed. This mechanism predicts the size, spacing and position of the clusters along the nanotube axis. Furthermore, this study shows that the 1D clusters display clear fingerprints of J-aggregation for bright and stable photon emission properties.

**Results and discussion**

The preparation of the samples begins with liquid-phase encapsulation of α-sexithiophene (6T) molecules inside pre-opened and cleaned BNNTs to form the so-called 6T@BNNT assembly (Figure 1a). A transparent film made of poly(methyl-methacrylate) (PMMA) solution loaded with individually dispersed 6T@BNNTs is then prepared and dried. The homogeneous mixing of the tubes with the polymer chains results in a random orientation of the 6T@BNNTs in the matrix (see Supplementary information file S3) The film is finally stretched at 150°C to induce a movement of BNNTs embedded in the host matrix using a typical film expansion of ~100%. (Figure 1b and see Methods and Supplementary Information for details). Another molecule, a derivative of 3,6-bis[2,2']bithiophenyl-5-yl-2,5- di-n-oc-tylpyrrolo[3,4-c]pyrrole-1,4-dione (DPP2), is also used to test the clustering effect observed after bending on another molecules. The latter sample is labeled DPP2@BNNT.

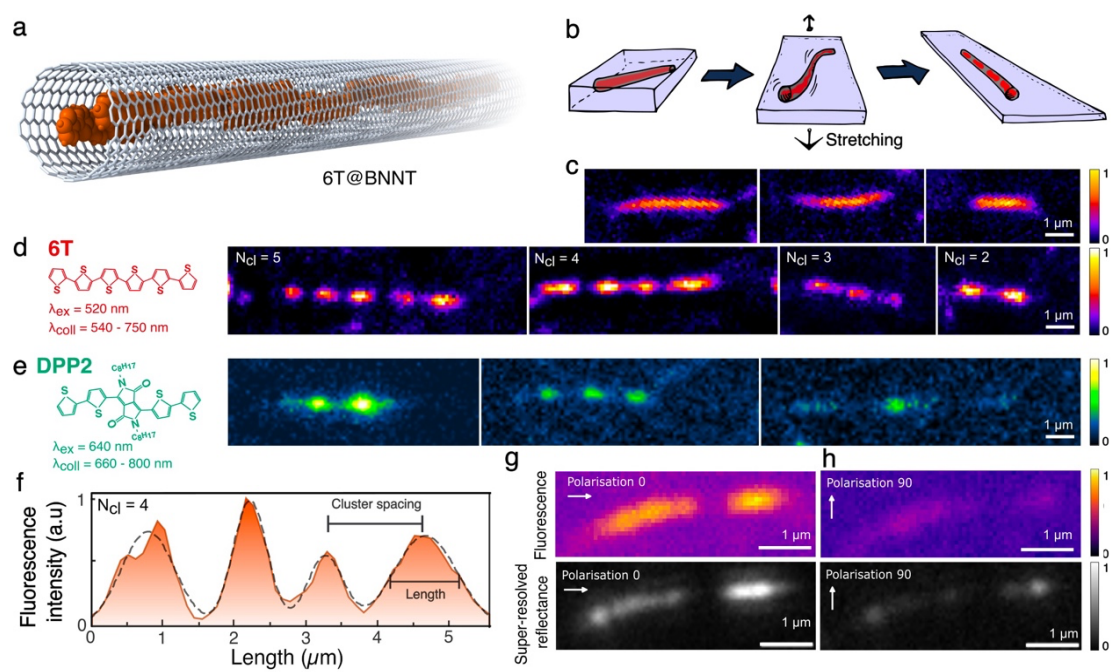

**Figure 1. Confocal fluorescence and super-resolved reflectance imaging of 6T@BNNTs and DPP2@BNNT. a** Schematic representation of a BNNT filled with 6T molecules. **b** Sketch of the experiment leading to molecular clustering. **c-e** Confocal fluorescence imaging after deconvolution of typical 6T@BNNTs (c,d) found in PMMA films before (c) and after (d) the stretching step at 150°C. The same was tested using and DPP2@BNNT (e). Left: drawing of the 6T and DPP2 molecules with indication of the excitation and collection conditions used for fluorescence imaging. **f** Fluorescence intensity profile and multi-peak gaussian fit (dashed black line) of a 6T@BNNT sample with 4 clusters ($N_{cl}=4$). The period, and the Full-Width Half-Maximum (FWHM) are used to statistically characterize the cluster morphologies. **g** Optically polarized fluorescence signals from the same 6T@BNNT (top) compared with super-resolved confocal reflectance imaging (down). **h** Polarisation dependency of the fluorescence and reflectance from the clustered 6T@BNNT presented in (g). The optical resolution in fluorescence images is of 250 nm after deconvolution.

Figures 1c,d show a set of fluorescence images of 6T@BNNTs localized in the PMMA film before and after stretching (See Methods and supplementary for imaging conditions). Before stretching (Fig. 1c), 6T@BNNTs exhibit a distribution of fluorescence signal from the 6T molecules all along the BNNT axis. The signal is uniform at the scale of the optical resolution of our setup (~220 nm). This result is consistent with previous work on uniform molecular packing along the nanotube length.[38] After stretching the PMMA film, a strong modulation of the fluorescence signal is detected, giving a regular sequence of bright and dark segments along the BNNT axis, as shown in Fig. 1d. This phenomenon has been clearly observed on many BNNTs of various lengths, between 1.5 and 15 µm (see Figure 1d and S3 for additional examples). Bright segments between 0.6 and 2 µm in size are found, which represents the equivalent of approximately 300 and 2000 molecules, respectively. The number of sequences varies from 2 to 10 and depend on the nanotube. Figure 1e also shows periodic modulations of the signal using DPP2 molecules, which shares the same rod-like shape as 6T but has a different axial structure. Therefore, this clustering effect is general and does not depend on the choice of rod-like molecules.

To determine whether this unexpected modulation originates from fluorescence quenching or local molecular depletion within the nanotube cavity, we performed correlative images from the same 6T@BNNT using two different modalities: confocal reflectance and fluorescence microscopy (technical details are given in Supplementary Information). While fluorescence is driven by the radiative recombination of excitons, contrast in reflectance originates instead from a change of refractive index, which implies a change in structure or of nature of matter. Both microscopy modalities in Figure 1g indicate a strong spatial correlation between dark segments. Knowing that empty BNNTs in PMMA display no or very weak fluorescence signal in reflectance, this comparison demonstrates that the molecules are most likely absent in the dark parts. We therefore rule out the hypothesis of a continuous encapsulated aggregates with periodic quenching of the luminescence along the BNNT axis. In addition, Figures 1g-h and Supplementary Information section 5 show that both signals are related to the light polarization with respect to the tube axis, which further indicates a strong molecular alignment inside the BNNTs.[30,39] Altogether, these observations point towards an efficient and long-range motion of molecules forming periodic patterns on the micrometre scale. From now on, these bright and dark segments are referred to clusters and gaps, respectively.

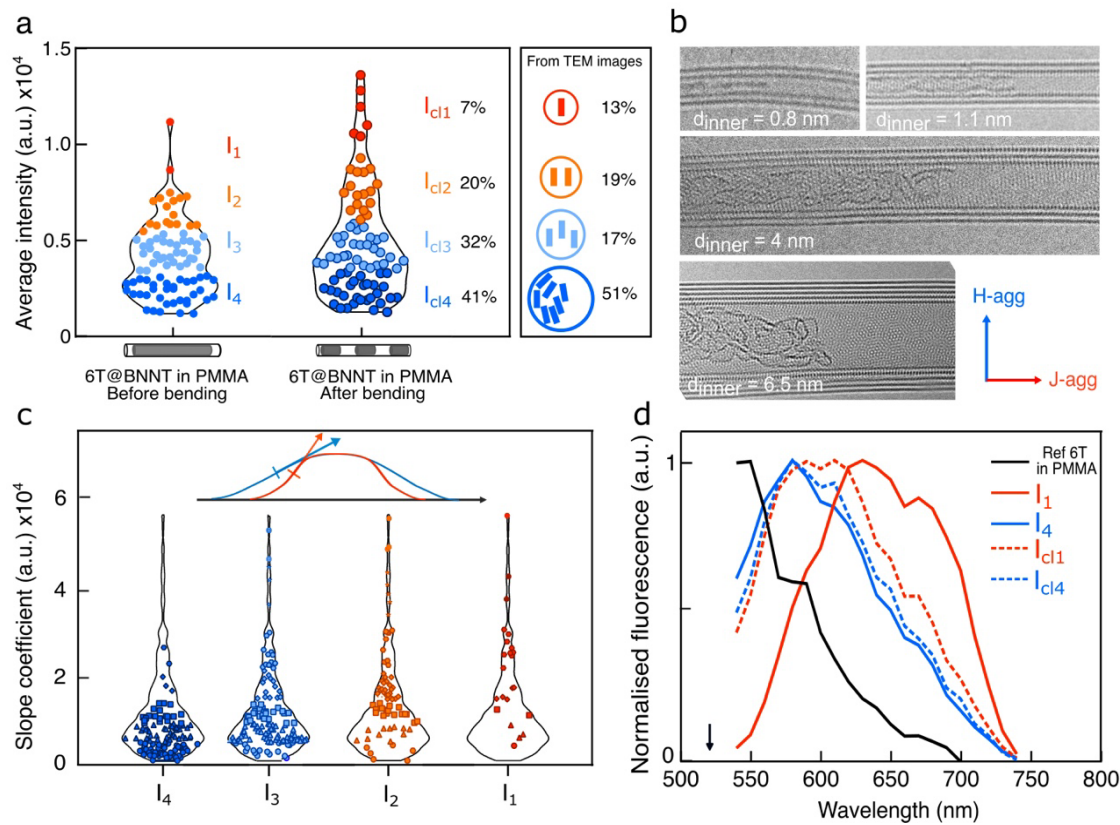

**Figure 2. Clustered 6T@BNNT: Emission enhancement in 1D J-aggregates with small BNNT diameters. a** Comparison of the average intensity distribution of 6T@BNNT fluorescence before and after deformation. The intensities $I_1, I_2, I_3, I_4$ and $Icl_1, Icl_2, Icl_3, Icl_4$ correspond to the average intensity of each subpopulation before and after deformation, respectively. Right panel inset: proportion of 6T@BNNT having *n* rows of molecules inside the tube cavity, adapted from ref[39]. **b** ac-HRTEM images of 6T@BNNTs of different internal diameters, adapted from ref[39] **c** Tangent slope at FWHM of the intensity profile for 320 clusters depending of their initial $I_1$, $I_2$, $I_3$, $I_4$. **d** Normalized emission spectra of free 6T in PMMA, 6T@BNNT in PMMA at average intensity $I_1$ and $I_4$, 6T@BNNT clustered in PMMA at average intensity $I_{cl1}$ and $I_{cl4}$ with an excitation laser at 520nm.

We first investigated the optical properties and aggregation states of the confined 6T molecules before and after mechanical stretching. Figure 2a presents the average integrated intensity of luminescent segments on non-clustered and clustered 6T@BNNTs, respectively. Surprisingly, the envelope of the distribution indicates a discretization of the 6T@BNNTs in sub-populations around specific intensities. A group differentiation study using K-mean and elbow analysis (see

Supplementary Information) shows that the intensity values fall in four sub-populations, as indicated in Figure 2a using color coding. For non-clustered and clustered 6T@BNNTs, the subpopulations, here labelled as $I_4,...,I_1$ and $I_{cl4},...,I_{cl1}$, are distributed on a similar range of intensities with a factor of 6 ±1 between the lower and larger intensity values. Interestingly, the percentage value for each sub-population is similar to those observed on the number of molecular rows included in a given BNNT section, which is directly dictated by the BNNT inner diameter[39]. This organization in rows can be visualized using an aberration-corrected High Resolution Transmission Electronic Microscope (ac-HRTEM). The images in Figure 2b show 6T@BNNTs having an heterogeneous filling factor and a lower ordering of the molecules at the ends of the aggregates for large inner diameters. In contrast, abrupt interfaces are observed in small inner diameters, which all have sharp empty/filled frontiers. Further, we plotted for each subpopulations, $I_4...I_1$, the slope of the cluster intensity at FWHM, as an indicator of the abruptness of empty/filled interfaces in 6T@BNNT (Figure 2c). A progressive increase of the slope from $I_4$ (lower intensity) to $I_1$ (higher intensity) is clearly seen. These results strengthen the link between intensity and their inner diameter of 6T@BNNTs.

It is counter-intuitive that 6T@BNNTs formed with one or two rows of 6T emitters appear in Figure 2 brighter than 6T@BNNTs made of 4 or more rows, but this difference can be ascribed to the aggregation state of the confined 6T molecules, which emissivity is directly linked to the morphology and packing order. For example, a face-to-face stacking of the molecules generates H-aggregates, which state is characterized by a weak emission and blue-shifted wavelength. This contrasts with a head-to-tail organization of the molecules, which results in very bright J-aggregation states and redshifted wavelengths. From the ac-HRTEM images of 6T@BNNTs in Figure 2b, we can see that differently weighted H- and J-aggregation states coexist in a given 6T@BNNT, each with different number of rows and degrees of molecular alignment. For example, small-diameter 6T@BNNTs have only one or two rows of 6Ts, and therefore the J-aggregation character should dominate when the inner space is a constraint. Conversely, in large inner diameters, H-aggregation can compete with J-aggregation, and this results in a lower emissivity. To test this hypothesis, we recorded the fluorescence spectra from 6T@BNNTs selected within the $I_4/I_{cl4}$ and $I_1/I_{cl1}$ sub-populations. (Figure 2d). The results first show that molecules confined in BNNTs, independently of the subpopulation, have an emission energy at least 50 nm redshifted from that of free molecules in PMMA (dark curve). This shift is a strong fingerprint of J-aggregation in 6T@BNNT and it is consistent with a good alignment of the molecules inside the BNNTs, as discussed in ref[38,39]. Furthermore, it can be noted that the brighter subpopulations $I_1/I_{cl1}$ exhibit an accentuated red shift compared to the less bright sub-populations $I_4/I_{cl4}$. This validates the hypothesis of a mixed HJ-aggregation in 6T@BNNT with reinforced J-aggregation for the smallest inner diameters BNNTs. Hence the densification of molecules induced by their diffusion into clusters enhances the intermolecular coupling within the aggregates, and leads to brighter fluorescence emission, especially for small diameters BNNTs.

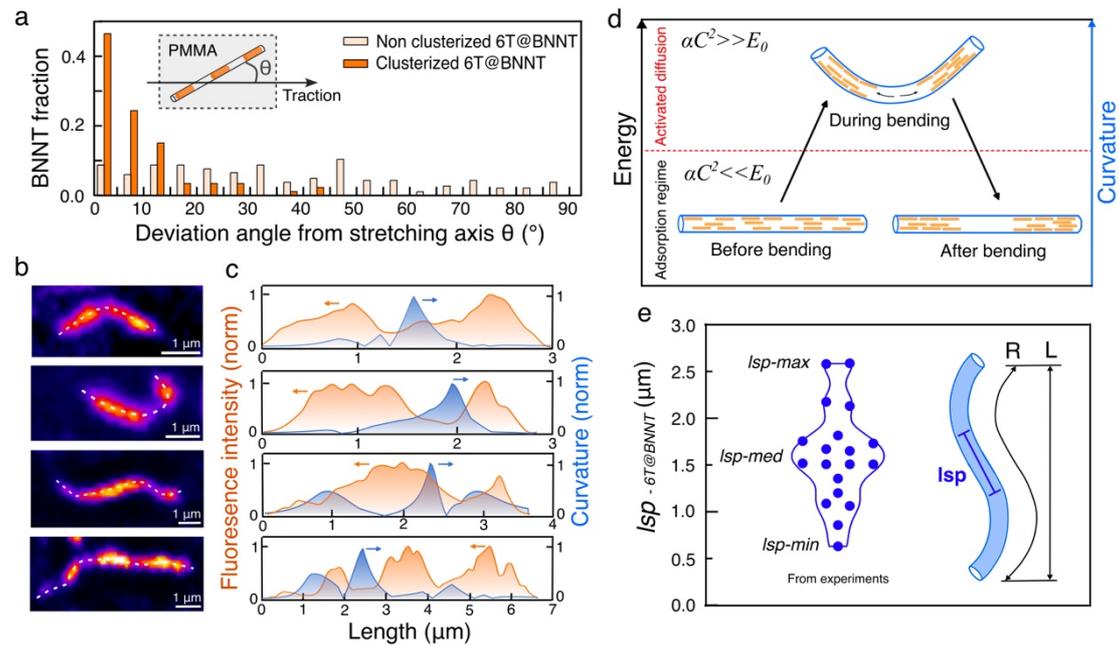

**Figure 3. Driving force of molecular active diffusion in 6T@BNNT. a** Distribution of clustered and un-clustered populations of 6T@BNNT as a function of the deviation angle from the traction axis. **b** Confocal fluorescence images of representative curved 6T@BNNTs recorded at the early stage of the PMMA/6T@BNNT stretching step. The dashed white lines highlight the tube axis along which intensity profiles are extracted. **c** Superposition of the intensity (in orange) and curvature profiles (in blue) of the 6T@BNNTs presented in (b) and highlighted by a dashed white line. **d** Potential energy diagram of the curvature-induced diffusion of 6T molecules inside BNNTs. **e** Distribution of the static bending persistence length measured from 19 curved 6T@BNNTs (left). Inset: A schematic representation of the static bending persistence length (*lsp*).

To elucidate the underlying mechanism of pattern formation, we analysed many individual 6T@BNNTs (N=85) detected in many 40x40x30 $\mu m^3$ confocal fluorescence datacubes acquired from different PMMA/6T@BNNT films. As shown in Figure 3a, a fraction of the 6T@BNNTs remain non-clustered. Interestingly, this sub-population corresponds to poorly oriented 6T@BNNTs in the film with respect to the film stretching direction. In contrast, almost 70% of the 6T@BNNTs with clusters after stretching were aligned within 10 degrees along the stretching axis.

To determine possible correlation between temperature and deformation of the matrix in the diffusion mechanism, we realized two additional experiments. In the first one, no stretch was applied during the annealing phase at 150°C and the results show no 6T@BNNT alignment or clustering (Figure S3). In the second experiment, we stretched an elastic matrix made of polydimethylsiloxane (PDMS) and 6T@BNNTs and found evidence of clustering when stretched at room temperature (Figure S3). These results clearly indicate that the displacement of the molecules inside the BNNT (i) is associated to a movement of the BNNTs embedded into the polymeric matrix, (ii) is temperature independent, and (iii) is not affected by the type of molecules and polymer matrix investigated.

To further strengthen the link between molecular migration inside the 6T@BNNTs and BNNTs bending within the PMMA matrix, we investigated the early stage of the stretching step by probing PMMA/6T@BNNTs films stretched by less than 3%. We also removed the film from the hot plate immediately after stretching, which readily freezes the film. Figure 3b shows representative confocal fluorescence images of 6T@BNNTs detected in these samples. First,

6T@BNNTs display curved or serpentine shapes, which drastically contrast with the straight rod shapes systematically found in 6T@BNNTs before and after full stretching. Second, the clusters are found to be already well-structured, indicating a fast long-range directed migration of thousands of molecules within only few seconds, albeit very confined within a 1D channel. More strikingly, overlaying in Figure 3c the BNNT curvature profiles with that of the intensity profiles of the 6T fluorescence, both extracted along the tubes axis highlighted in Figure 3b, reveals a clear correlation between the cluster positions, gaps, and BNNT deformations. The curvature of the 6T@BNNTs was extracted with the Kappa plugin on FIJI (more details are available in the Supplementary Information file).[40] Indeed, the curvature maxima along the BNNT coincide with either the beginning or the end of a cluster. As a corollary, the intensity maxima of the clusters are mainly located in the straight parts of the BNNTs. Combined with the large aspect ratio of the cluster, i.e micron scale in length and nanoscale in diameter, the trend enables us to rule out formation mechanisms such as Ostwald ripening, Plateau-Rayleigh instabilities, wetting effects, and quasi 1D spinodal phase transition. Indeed these mechanisms applied on solid or liquid tubular architectures and usually lead to the formation of clusters with size and spacing related to the scale of the diameter of the tubule.[41–44] It should correspond in our case to clusters of 1-10 nm in length, which is not the case. The size and periodicity of the micron-scale patterns can also be used to rule out the hypothesis of molecule migration induced by BNNT buckling and pinching, as this can only occur during bending in BNNT segments of the order of a few dozen of nanometers.[45–47]

The results support, however, an alternative mechanism in which there is activated diffusion of the molecules in the bent part of the BNNTs. A modulation of the molecular affinity with longitudinal curvature can explain both the guided diffusion of the molecules and the break of symmetry along the BNNT axis. That is, a strong 1D confinement is induced by the BNNT wall and this occurs over a cross-section of the order of a few van der Waals distances (4 - 10 Å), giving radial symmetry of van der Waals forces. The confinement produces competing forces on the molecules and reduces the exerted adsorption force between the molecules and the wall. In addition, the crystalline nature of the BNNT walls ensures an energy equivalence between two adjacent adsorption sites at the single-molecule scale, reducing the cost of the elementary motion step. On one hand, ease of movement is evidenced by the size of the gaps observed, which is in the micrometer range, and therefore molecules can migrate over distances corresponding to 1000 times their own size. Consequently, the energetical barrier between adsorption and diffusion of molecules in BNNT must be significantly lowered in the process. On the other hand, the elongated molecules used in this study have an axis of rotational symmetry that coincides with the longitudinal axis of the BNNT. The frustration of this axis of symmetry by the local curvature of the BNNT statistically induces by energy minimization the migration of molecules towards non-curved portions of the BNNT.

Following these considerations, we propose a phenomenological model of molecular transport in BNNTs; its central microscopic hypothesis is that the adsorption energy, $E$, of molecules on the BNNTs' wall depends on the local curvature, as highlighted in Figure 3d. This can be written minimally as $E=\alpha C^2 - E_0$, where $\alpha > 0$ is a phenomenological constant which depends on $C$ the local curvature of the BNNT, and $E_0 > 0$ the adsorption energy for straight BNNTs. The unbinding rate of molecules from the walls can then be written as $K=K_0 e^{\beta E}$, where $\beta$ is the inverse temperature. This leads to the following diffusion equation for the concentration $c(x,t)$ of molecules along a BNNT.

$$\partial_t c(x,t) = \partial_x^2 (D(x)c(x,t)) \quad \text{Eq. (0a)}$$

, where $x$ denotes the coordinate along the BNNT. The diffusion coefficient, $D(x)$, depends on curvature according to $D(x) \propto K(x)$.[48] The observed stability of patterns for straight tubes after the stretching phase suggests that $\beta E_0 >> 1$. This is consistent with a vanishing $D$ for straight tubes, and thus an arrested dynamics of molecules migration in this phase. However, during the stretching phase, the local strain of the polymer matrix triggers random deformations of BNNTs, and domains with $C \gtrsim \sqrt{E_0/\alpha}$ become activated with non-vanishing $D$. For a given deformed configuration during the stretching phase and assuming the deformation dynamics of BNNTs due to stretching slower than the relaxation of *c(x)*, the stationary concentration profile in activated domains satisfies:

$$c(x) \propto \frac{1}{D(x)}) \propto \frac{1}{K(x)} \propto e^{-\beta \alpha C^2} \qquad \text{Eq. (0b)}$$

This argument shows that the concentration of molecules is minimized in BNNT domains with maximal curvature, and is thus maximized in straight domains, as observed experimentally.

We characterized the straight domains by using the static persistence length (*lsp*), which accounts for the maximum straight length in a filament curved by a static deformation (see Figure 3e). Based on the work by Lee et al, [49,50] the quantity *lsp* can be derived from the spatial average of the square end-to-end vector $\langle R^2 \rangle$ and the bending ratio $D_b$ of a rigid random rod coil, as described in Equations Eq. S2 and 1a:

$$D_b \equiv \frac{\langle R^2 \rangle}{L^2} \simeq \left(\frac{2lp_0}{L}\right)\left(\frac{1+\cos(\theta)}{1-\cos(\theta)}\right) = M\left(\frac{2lp_0}{L}\right) = \frac{2lsp}{L} \qquad \text{(Eq. 1a)}$$

, where $lsp = M lp_0$ is the static bending persistence length, $lp_0$ is a straight segment length and *M* is a constant for a fixed bending angle of a BNNT with a given stiffness. Hence, we experimentally determined the distribution of *lsp* values from a statistical set of 19 curved 6T@BNNTs, as presented in Figure 3e. From this distribution, we define $lsp_{-min}$ = 630 nm, $lsp_{-max}$ = 2.6 µm and $lsp_{-med}$ = 1.5 µm, as being the extrema and median values of *lsp*. The *lsp* values are well below the equilibrium persistence length [49] of BNNTs, which is estimated at several hundred of microns due to their very high Young's modulus[51–54] (see Supplementary Information for details). Hence, the bending of 6T@BNNTs observed in the PMMA matrix originates from heterogeneous local deformations of the matrix during stretching and not from thermally induced fluctuations during the annealing at 150°C. This also explains the absence of cluster in the PMMA/6T@BNNT films annealed without stretching.

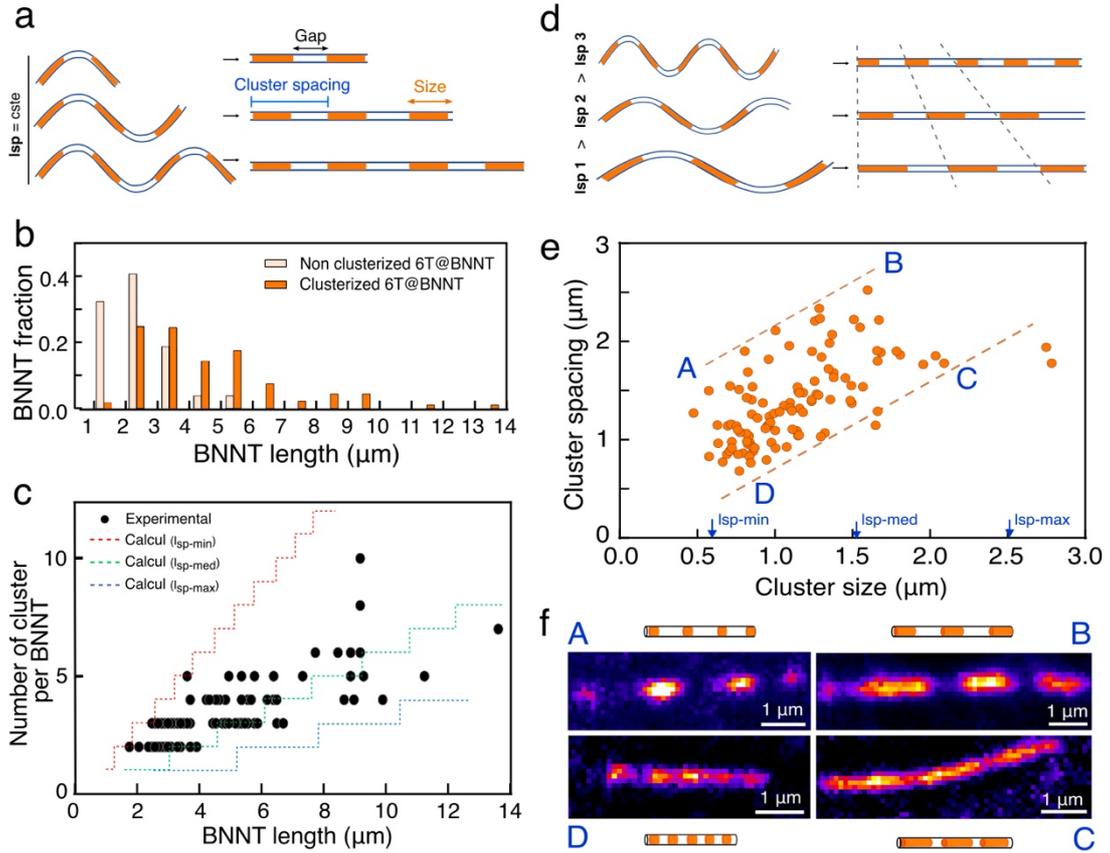

**Figure 4. Morphology and patterns of clusters in 6T@BNNTs. a** Schematic representation of a clustered 6T@BNNTs from 6T@BNNTs having constant *lsp* and different lengths. **b** Distribution of clustered and non-clustered 6T@BNNT population as a function of the BNNTs length. **c** Calculated and experimental measurements of the number of clusters per 6T@BNNT as a function of BNNTs length. **d** Schematic representation of a clustered 6T@BNNT from 6T@BNNTs having constant length and different *lsp* values. **e** Mean cluster spacing in a 6T@BNNT as a function of their mean cluster size. **f** Selected examples of clustered 6T@BNNTs with different sizes and periods, corresponding to the regions A, B, C, and D highlighted in (e).

To further validate the hypothesis of an activated diffusion induced by a longitudinal bending in BNNTs, we compare in Figure 4 the characteristic size and distribution of the clusters according to the bending properties and length of 6T@BNNTs on a set of 320 clusters from 85 individual 6T@BNNTs. In Figure 4c, we examined the number of clusters ($N_{cl}$) per 6T@BNNT as a function of BNNT length. It appears first that a minimal threshold value of 1.5±0.3 μm for the BNNT length is necessary to allow cluster formation. As illustrated in Figure 4a, this value matches twice the *lsp-min* added by one gap length. Furthermore, the experimental dependency of $N_{cl}$ on the BNNT length in Figure 4c indicates that going from *n* to *n+1* clusters implies an additional BNNT length corresponding to *lsp* in 6T@BNNT. It can be seen that all experimental values of $N_{cl}$ as a function of BNNT length are bounded by the staircase lines simulated from the experimental values of *lsp-min*, *lsp-max* and *lsp-med* shown in Figure 3e, (More details on the calculation of the number of clusters as a function of BNNT length are available in the Supplementary Information file).

Other significative hallmarks are presented in Figure 4e, which highlights the relationship between cluster spacing and size within a given 6T@BNNT. First, it can be seen that the smallest and the largest cluster lengths measured corresponds to *lsp-min* and *lsp-max*, respectively. Second, a strong correlation is observed between size and spacing of cluster within a given 6T@BNNT. This is consistent with our diffusion model because an increase of *lsp* (size of the cluster) induces automatically larger spacing between two curvature maxima, as illustrated in Figure 4d. This also homothetically implies a lengthening of the clusters. These results demonstrate the close link between the mechanical bending properties of a BNNT and the morphology of the clusters, which is determined by construction by the length and *lsp*. To illustrate these trends, Figure 4f presents the four distribution extrema of clusters, which correspond to clustered 6T@BNNT selected in Figure 3d in the areas labelled A, B, C and D. These images underlines that all the combinations size/spacing in between these four cases are obtained by playing with *lsp*. These results open up new perspective for the manipulation and positioning of molecules on the nanoscale and the formation of 1D aggregates of finite and controlled size.

In summary, we have shown how a BNNT can be used as a template to infer controlled diffusion of molecules confined into the 1D cavity of the nanotube. Our study reveals an efficient molecular diffusion is activated in the bended part of the BNNTs under mechanical stimuli. Using energetic arguments, we described the guided diffusion phenomenon using a simple model based on molecular adsorption and diffusion regimes in the BNNT. The results also reveal that the bending curvature gradient determines the direction of the molecular migration along the BNNT axis. This new diffusion mechanism enables us to obtain periodic formation of bright and aligned J-aggregates with micrometer periodic spacings and finite lengths. This finding paves the way to a novel deterministic method to elaborate supramolecular architectures with nanotubes and highlights how ordered patterns of molecules can be formed at the nanoscale. This method can be leveraged to explore various applications in photonics.

**Methods :**

**Materials**. Boron nitride nanotubes (BNNT) were provided by BNNT LLC and the Canadian National Research Center (NRC). Only reagent grade solvents were used, α-sexithiophene (6T), DPP2 was obtain by oligothiophene derivatives, such as 3,6-Bis-[2,2']bithiophenyl-5-yl-2,5-di-n-octylpyrrolo[3,4-c]pyrrole-1,4-dione (DPPx), were synthesized according to the procedure in Reference[55]. Poly(methyl methacrylate) (PMMA, Mw=120000 g.mol-1) were purchased from Sigma-Aldrich and used as received, SYLGARD™ 184 Silicone Elastomer were purchased from Neyco and used as received. See Section S2.1, Supporting Information, for details on cleaning and encapsulation of BNNTs.

**Fabrication and stretching of the PMMA- or PDMS-mol@BNNT films**. An amount of 0.33g of PMMA (Mw=120000 g.mol$^{-1}$) was dissolved in 10 mL of DMF with a magnetic stirrer during 4 hours. Then 1 mL from the Mol@BNNT solution were added before drying in a petri dish. (see section S3.1 for more details.)

The film were then installed in the traction system for stretching, as presented in figure S3. The starting distance between the two jaws is 5 mm. The heating plate is heated to a temperature of 150°C. The system is heated for 15 min. The stretching is then carried out by stretching the film using the translation stage at a speed of 1mm/min roughly. The distance between the two edges is then measured to quantify the stretching. Depending of the films, they are then stretched over 150% at 150°C in 20 mins by steps of 3% to induce the movement of the BNNTs in the host matrix.

**Confocal fluorescence imaging**. The data presented were acquired with a Leica microscope SP8 WLL2 on an inverted stand DMI6000 (Leica Microsystems, Mannheim, Germany) using a HCX Plan Apo CS2 63X oil NA 1.40 objective.

For the deconvolution of the images obtained from the microscope, PMMA films loaded with fluorescent fluo-sphere 0.170 µm in diameter (Azide 2mM) at 524,532,540 nm laser excitation and 560-640 nm range collection.

**Super-resolved reflectance imaging**. The reflectance images of 6T@BNNTs in Figure 1g and 1h were acquired using a confocal-based microscopy where the Image Scanning Microscopy (ISM) technique is implemented to achieve high speed label-free super-resolved imaging. (Figure S.4) The non-polarized white beam from a super continuum laser is spread into chromatic line by a diffraction grating. The chromatic line is projected to the sample by an objective (OBJ, 60X, water immersion, NA=1.27) and scanned by a resonant mirror (RM, @12kHz). The back-scattered signal is collected by the same objective and travel back to the diffraction grating. A pinhole is placed in a conjugated image plane to reject the background signal. The reflected signal is spread again by second a diffraction grating and rescanned by the other face of the resonant mirror. A polarisation sensitive camera (Alkeria, Celera C5S-MP) is used to acquire images of different polarization states at the same time.

**Numerical analysis of the clusters.** The characterization of of the static persistence length *lsp* is presented in the Supporting Information section S.6.2.

To quantify the intensity subpopulations of 6T@BNNTs presented in figure 4, we applied a K-mean analysis Using the sklearn.cluster. KMeans library on Python with the following parameters: model = KMeans(n_clusters = n, init='k-means++', max_iter=500, random_state=42). (n varying from 1 to 10). By fitting with the average intensities, we obtain the inertia value for each groupe n available in figure S.7 in the SI. Using the KElbowVisualizer function in the Yellowbrick library, we can locate the number of groups for which the inertia starts to decrease linearly. The inertia is calculated by measuring the distance between each data point and its centroid, squaring this distance, and summing these squares across one group. We also present a Silhouette Coefficient representation for the same populations.

The characterization of the slope coefficient of the tangent at the FWHM abscissa for each cluster presented in Figure 4c were carried out with Pytho is detailed in Supporting Information Figure S9.

**Acknowledgments**

The authors warmly acknowledge B. Simard and his team at the National Research Council Canada for donation of BNNTs materials. E.G. acknowledges funding from CNRS starting package funding and CNRS Tremplin program, the GDRi Graphene and Co, the GDRi multifunctional nano for travel support. E. G. and G. R. are supported by the CNRS MITI 'Défi Auto-Organisation' grant. G. R. acknowledges the GdR ImaBio for support, and the ANR for funding (ANR-21-CE45-0028). R.M. was supported by Discovery grants RGPIN-2019-06545 and RGPAS-2019-00050 (R.M.), by the Canada Research Chair and the Canada Foundation for Innovation (FCI). Confocal fluorescence microscopy was performed on the Bordeaux Imaging Center a service unit of the CNRS-INSERM and Bordeaux University, member of the national infrastructure France BioImaging supported by the French National Research Agency (ANR-10-INBS-04). P.B. acknowledges the European Research Council (ERC) under the European Union's Horizon 2020 Research and Innovation Programme (grant agreement No. 848645).

**Authors Contributions**

J.-BM, G.R., E.G. designed the experiments, J.-BM prepared the samples, J.-BM, D.-MT, A.A., P.B., A.L., R.M., G.R, E.G. performed the experiments and/or analyzed the results. J.-BM, EG, R.V. discussed the diffusion mechanism. E.G. supervised the work. All authors contributed to the scientific discussions, manuscript preparation and final version.

**Competing Interest Statement**

The authors declare no competing financial interests.

**References**

[1]     F. C. Spano, C. Silva, *Annu. Rev. Phys. Chem.* **2014**, *65*, 477.
[2]     H. Manzano, I. Esnal, T. Marqués-Matesanz, J. Bañuelos, I. López-Arbeloa, M. J. Ortiz, L. Cerdán, A. Costela, I. García-Moreno, J. L. Chiara, *Adv. Funct. Mater.* **2016**, *26*, 2756.
[3]     J.-B. Trebbia, Q. Deplano, P. Tamarat, B. Lounis, *Nat Commun* **2022**, *13*, 2962.
[4]     G. Rainò, M. A. Becker, M. I. Bodnarchuk, R. F. Mahrt, M. V. Kovalenko, T. Stöferle, *Nature* **2018**, *563*, 671.
[5]     G. Rainò, H. Utzat, M. G. Bawendi, M. V. Kovalenko, *MRS Bulletin* **2020**, *45*, 841.
[6]     A. Eisfeld, C. Marquardt, A. Paulheim, M. Sokolowski, *Phys. Rev. Lett.* **2017**, *119*, 097402.
[7]     H. Ambronn, *Ber Deutsch Botan Ges* **1888**, 85.
[8]     H. Jablosky, *Nature* **1934**, *133*.
[9]     E. W. Thulstrup, J. Michl, *Spectrochimica Acta Part A: Molecular Spectroscopy* **1988**, *44*, 767.
[10]    G. Rainò, T. Stöferle, C. Park, H.-C. Kim, I.-J. Chin, R. D. Miller, R. F. Mahrt, *Advanced Materials* **2010**, *22*, 3681.

[11]     M. Washizu, S. Suzuki, O. Kurosawa, T. Nishizaka, T. Shinohara, *IEEE Transactions on Industry Applications* **1994**, *30*, 835.
[12]     I. V. Fedorov, I. I. Bobrinetskiy, B. I. Shapiro, A. V. Romashkin, V. K. Nevolin, *Physics Letters A* **2014**, *378*, 226.
[13]     K. J. Modica, Y. Xi, S. C. Takatori, *Front. Phys.* **2022**, *10*, 869175.
[14]     J. Prost, F. Jülicher, J.-F. Joanny, *Nature Phys* **2015**, *11*, 111.
[15]     J. Palacci, B. Abécassis, C. Cottin-Bizonne, C. Ybert, L. Bocquet, *Phys. Rev. Lett.* **2010**, *104*, 138302.
[16]     A. Walther, A. H. E. Müller, *Chem. Rev.* **2013**, *113*, 5194.
[17]     A. Ashkin, J. M. Dziedzic, J. E. Bjorkholm, S. Chu, *Opt. Lett., OL* **1986**, *11*, 288.
[18]     M. Veiga-Gutiérrez, M. Woerdemann, E. Prasetyanto, C. Denz, L. De Cola, *Advanced Materials* **2012**, *24*, 5199.
[19]     C. J. Bustamante, Y. R. Chemla, S. Liu, M. D. Wang, *Nat Rev Methods Primers* **2021**, *1*, 1.
[20]     R. Mittal, E. Karhu, J. Wang, S. Delgado, R. Zukerman, J. Mittal, V. M. Jhaveri, *Journal Cellular Physiology* **2019**, *234*, 1130.
[21]     A. Rustom, R. Saffrich, I. Markovic, P. Walther, H.-H. Gerdes, *Science* **2004**, *303*, 1007.
[22]     R. D. Vale, *Cell* **2003**, *112*, 467.
[23]     S. Klumpp, R. Lipowsky, *Proceedings of the National Academy of Sciences* **2005**, *102*, 17284.
[24]     R. Van Hameren, P. Schön, A. M. Van Buul, J. Hoogboom, S. V. Lazarenko, J. W. Gerritsen, H. Engelkamp, P. C. M. Christianen, H. A. Heus, J. C. Maan, T. Rasing, S. Speller, A. E. Rowan, J. A. A. W. Elemans, R. J. M. Nolte, *Science* **2006**, *314*, 1433.
[25]     H. Tanaka, T. Sigehuzi, *Phys. Rev. Lett.* **1995**, *75*, 874.
[26]     D. M. Agra-Kooijman, G. Singh, A. Lorenz, P. J. Collings, H.-S. Kitzerow, S. Kumar, *Phys. Rev. E* **2014**, *89*, 062504.
[27]     H. He, H. Li, Y. Cui, G. Qian, *Advanced Optical Materials* **2019**, *7*, 1900077.
[28]     M. Busby, C. Blum, M. Tibben, S. Fibikar, G. Calzaferri, V. Subramaniam, L. De Cola, *J. Am. Chem. Soc.* **2008**, *130*, 10970.
[29]     A. Kuzyk, R. Jungmann, G. P. Acuna, N. Liu, *ACS Photonics* **2018**, *5*, 1151.
[30]     E. Gaufrès, N. Y.-W. Tang, F. Lapointe, J. Cabana, M.-A. Nadon, N. Cottenye, F. Raymond, T. Szkopek, R. Martel, *Nature Photon* **2014**, *8*, 72.
[31]     S. Cambré, J. Campo, C. Beirnaert, C. Verlackt, P. Cool, W. Wenseleers, *Nature Nanotech* **2015**, *10*, 248.
[32]     J. Campo, S. Cambré, B. Botka, J. Obrzut, W. Wenseleers, J. A. Fagan, *ACS Nano* **2021**, *15*, 2301.
[33]     Y. Almadori, G. Delport, R. Chambard, L. Orcin-Chaix, A. C. Selvati, N. Izard, A. Belhboub, R. Aznar, B. Jousselme, S. Campidelli, P. Hermet, R. Le Parc, T. Saito, Y. Sato, K. Suenaga, P. Puech, J. S. Lauret, G. Cassabois, J.-L. Bantignies, L. Alvarez, *Carbon* **2019**, *149*, 772.
[34]     R. Chambard, J. C. Moreno-López, P. Hermet, Y. Sato, K. Suenaga, T. Pichler, B. Jousselme, R. Aznar, J.-L. Bantignies, N. Izard, L. Alvarez, *Carbon* **2022**, *186*, 423.
[35]     E. Gaufrès, N. Y.-W. Tang, A. Favron, C. Allard, F. Lapointe, V. Jourdain, S. Tahir, C.-N. Brosseau, R. Leonelli, R. Martel, *ACS Nano* **2016**, *10*, 10220.
[36]     A. Cadena, B. Botka, E. Székely, K. Kamarás, *physica status solidi (b)* **n.d.**, *n/a*, 2000314.

[37]   S. Wasserroth, S. Heeg, N. S. Mueller, P. Kusch, U. Hübner, E. Gaufrès, N. Y.-W. Tang, R. Martel, A. Vijayaraghavan, S. Reich, *J. Phys. Chem. C* **2019**, *123*, 10578.
[38]   C. Allard, L. Schué, F. Fossard, G. Recher, R. Nascimento, E. Flahaut, A. Loiseau, P. Desjardins, R. Martel, E. Gaufrès, *Advanced Materials* **2020**, *32*, 2001429.
[39]   A. Badon, J.-B. Marceau, C. Allard, F. Fossard, A. Loiseau, L. Cognet, E. Flahaut, G. Recher, N. Izard, R. Martel, E. Gaufrès, *Mater. Horiz.* **2023**, *10*, 983.
[40]   H. Mary, G. J. Brouhard, **2019**.
[41]   Y. Qin, S.-M. Lee, A. Pan, U. Gösele, M. Knez, *Nano Lett.* **2008**, *8*, 114.
[42]   L. Zhu, G. Lu, S. Mao, J. Chen, D. A. Dikin, X. Chen, R. S. Ruoff, **2007**.
[43]   M. E. Toimil Molares, A. G. Balogh, T. W. Cornelius, R. Neumann, C. Trautmann, *Applied Physics Letters* **2004**, *85*, 5337.
[44]   J.-T. Chen, T.-H. Wei, C.-W. Chang, H.-W. Ko, C.-W. Chu, M.-H. Chi, C.-C. Tsai, *Macromolecules* **2014**, *47*, 5227.
[45]   H. M. Ghassemi, C. H. Lee, Y. K. Yap, R. S. Yassar, *Nanotechnology* **2011**, *22*, 115702.
[46]   D. Golberg, P. M. F. J. Costa, O. Lourie, M. Mitome, X. Bai, K. Kurashima, C. Zhi, C. Tang, Y. Bando, *Nano Lett.* **2007**, *7*, 2146.
[47]   Y. Huang, J. Lin, J. Zou, M.-S. Wang, K. Faerstein, C. Tang, Y. Bando, D. Golberg, *Nanoscale* **2013**, *5*, 4840.
[48]   J.-P. Bouchaud, A. Georges, *Physics Reports* **1990**, *195*, 127.
[49]   P.-G. de Gennes, *Scaling Concepts in Polymer Physics*, Cornell University Press, **1979**.
[50]   H. S. Lee, C. H. Yun, H. M. Kim, C. J. Lee, *J. Phys. Chem. C* **2007**, *111*, 18882.
[51]   A. E. Tanur, J. Wang, A. L. M. Reddy, D. N. Lamont, Y. K. Yap, G. C. Walker, *J. Phys. Chem. B* **2013**, *117*, 4618.
[52]   N. G. Chopra, A. Zettl, *Solid State Communications* **1998**, *105*, 297.
[53]   V. K. Choyal, V. Choyal, S. Nevhal, A. Bergaley, S. I. Kundalwal, *Materials Today: Proceedings* **2020**, *26*, 1.
[54]   N. Fakhri, D. A. Tsyboulski, L. Cognet, R. B. Weisman, M. Pasquali, *Proceedings of the National Academy of Sciences* **2009**, *106*, 14219.
[55]   A. B. Tamayo, M. Tantiwiwat, B. Walker, T.-Q. Nguyen, *J. Phys. Chem. C* **2008**, *112*, 15543.

Supporting Information file for:

# Forming 1D Periodic J-aggregates by Mechanical Bending of BNNTs: Evidence of Activated Molecular Diffusion

J.-B. Marceau[1], D.-M Ta[2] A. Aguilar[2], A. Loiseau[3], R. Martel[4], P. Bon[2], R. Voituriez[5], G. Recher[1], and E. Gaufrès[1*]

[1] Laboratoire Photonique Numérique et Nanosciences, Institut d'Optique, CNRS UMR5298, Université de Bordeaux, F-33400 Talence, France
[2] XLIM UMR CNRS 7252, Limoges, France
[3] Laboratoire d'Étude des Microstructures, ONERA-CNRS, UMR104, Université Paris-Saclay, BP 72, 92322 Châtillon Cedex, France
[4] Département de chimie et Institut Courtois, Université de Montréal, Montréal, Québec H3C 3J7, Canada
[5] Laboratoire Jean Perrin et Laboratoire de Physique Théorique de la Matière Condensée CNRS Sorbonne Université, 4 place Jussieu, 75005 Paris, France

*Correspondence: etienne.gaufres@cnrs.fr

## Contents

### 1. Materials

Boron nitride nanotubes (BNNT) were provided by BNNT LLC and the Canadian National Research Center (NRC). Only reagent grade solvents were used, α-sexithiophene (6T), DPP2 was obtain by Oligothiophene derivatives, such as 3,6-Bis-[2,2']bithiophenyl-5-yl-2,5-di-n-octylpyrrolo[3,4-c]pyrrole-1,4-dione (DPPx), were synthesized according to the procedure in Reference[1]. Poly(methyl methacrylate) (PMMA, Mw=120000 g.mol-1) were purchased from Sigma-Aldrich and used as received, SYLGARD™ 184 Silicone Elastomer were purchased from Neyco and used as received.

### 2. Methods

**2.1- Encapsulation of 6T@BNNT,DPP2@BNNT**

The BNNT powder was first annealed at 800° in air for 2 hours, then sonicated in DMF using a cup sonicator until complete dispersion was observed. The solution was then centrifuged at 12,000g and the top half of the centrifuge tube was collected.

The purified BNNTs were dispersed in DMF or toluene at a concentration of 0.1 mg/mL.

The encapsulation was done in a round-bottom flask equipped with a condenser, under reflux. The concentration of the 6T, DPP2 solution was fixed at 5x10-6 M on toluene. Purified BNNTs were then introduced in the round bottom flask and the encapsulation was carried out for 48h at 85°C.

**2.2- Encapsulation of 6T in BNNT for TEM grid (Mo/SiO2)**

About 20μL of BNNT solution in DMF was drop-casted on a Molybdenum grid with SiO2 membrane decorated with holes. Before encapsulation, the TEM grid was annealed under vacuum at 800°C for two hours. The grid was inserted in the 6T encapsulation solution at 85°C for 6 hours. Following the encapsulation, the grid was rinsed for a few seconds in DMF, cleaned using an oxygen plasma (100W, 10 minutes) and a piranha treatment (2 minutes) to completely remove the excess of non-encapsulated dyes[2].

3. **Fabrication and stretching of the PMMA-6T films, PMMA-6T@BNNT films and PMMA-DPP2@BNNT films, PDMS-6T@BNNT films.**

An amount of 0.33g of PMMA (Mw=120000 g.mol-1) was dissolved in 10 mL of DMF with a magnetic stirrer during 4 hours for each synthesis of film, then depending of the composite wanted:

**3.1- PMMA-6T films**:
100 µL of a 6T solution in DMF at 5x10-6 M were added to the PMMA solution and were stirred during 1h.

**3.2- PMMA-6T@BNNT films:**
1 mL of a centrifugated 6T@BNNT solution in DMF were added to the PMMA solution and were stirred during 1h.

**3.3- PMMA-DPP2@BNNT:**
1mL of a centrifugated DPP2@BNNT solution in DMF were added to the PMMA solution and were stirred during 1h.

The solution was then poured into a 6 cm diameter glass petri dish. The petri dish was placed on a hot plate at 40°C to accelerate the evaporation of DMF for 4 hours. Samples of dimensions 5x15mm were cut with scissors in the films. They were then installed in the traction system presented in figure S3. The starting distance between the two jaws is 5 mm. The heating plate is heated to a temperature of 150°C. The system is heated for 15 min. The stretching is then carried out by stretching the film using the translation stage at a speed of 1mm/min roughly. The distance between the two edges is then measured to quantify the stretching.

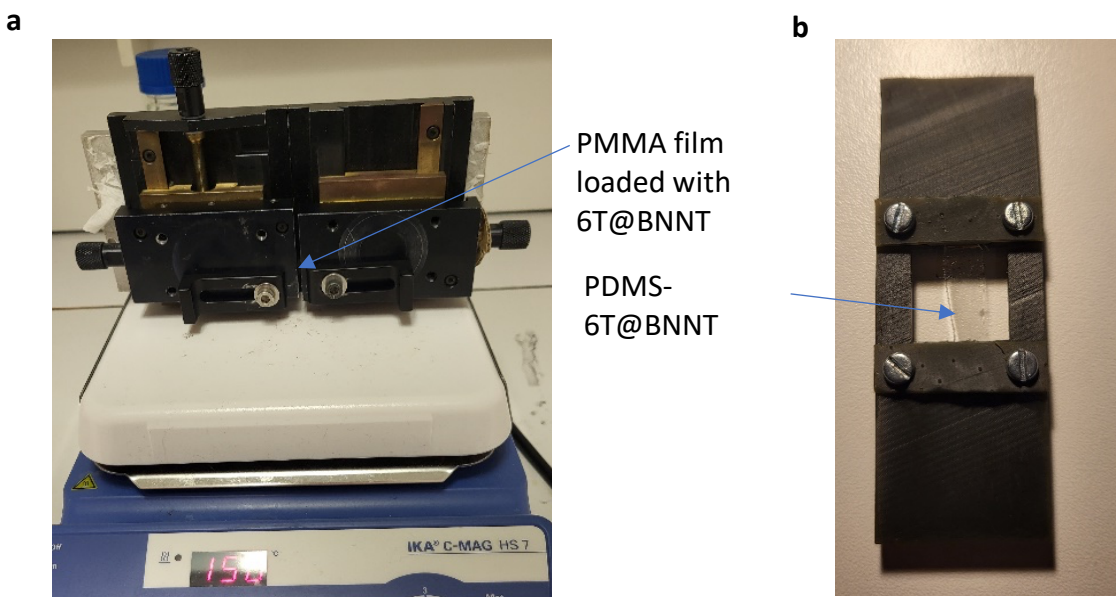

**Figure S1. (a) Custom-built stretching machine, based on optical translation stages, installed on a hot plate. The films are placed at a distance of 10 mm of the hot plate. (b) Sample carrier system for PDMS stretching at RT.**

**3.4- PDMS-6T@BNNT film:**

20g base agent of PDMS is first mixed with 2g of curing agent. 1mL of a centrifuged 6T@BNNT solution in DMF were added and the solution were stirred during 5 min with a glass stick. The solution was placed in a desiccator under vacuum during 30min for degassing. Then, the solution was poured into a 6 cm diameter glass petri dish. It was placed on a hot plate at 60°C to cure PDMS and accelerate the evaporation of DMF for 2 hours. Samples of dimensions 5x15mm were cut with scissors in the films. They were then installed in the traction system presented in figure S3. The starting distance between the two jaws is 5 mm. The stretching is then carried out by stretching the film using the translation stage at a speed of 1mm/min roughly. The PDMS film is either fixed using a 3D printed sample carrier (figure S3), before clamping screw removing, or just pulled out of the traction system for relaxation.

### 4. Confocal fluorescence Imaging PMMA-6T films, PMMA-6T@BNNT films, PMMA-DPP2@BNNT films and analysis

**4.1- Preparation of samples for confocal microscopy:**

To ensure appropriate immobilization and optimized optical collection under the microscope, the films were covered by refractive index liquid and sealed between a microscope slide and a glass cover slide.

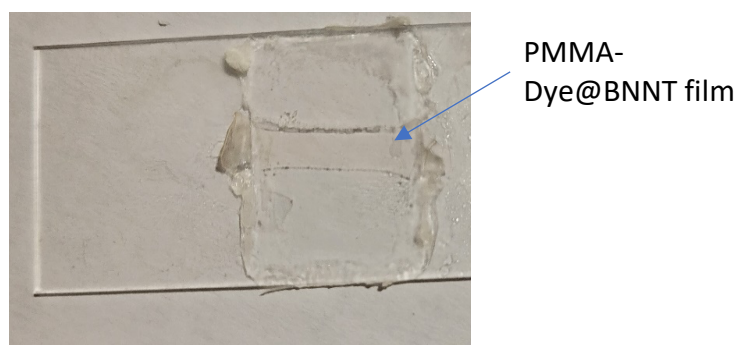

PMMA-Dye@BNNT film

**Figure S2. Immobilization of the film samples prior confocal imaging.**

**4.2- Confocal imaging parameters:**

The confocal fluorescence images data presented in Figure 1 (main text) were acquired with a Leica microscope SP8 WLL2 on an inverted stand DMI6000 (Leica Microsystems, Mannheim, Germany) using a HCX Plan Apo CS2 63X oil NA 1.40 objective. The excitation wavelength 6T@BNNTs and DPP2@BNNT are 510 nm and 660 nm, respectively. The spectral interval of collection for 6T@BNNTs and DPP2@BNNT are 570-750 nm and 680-794 nm, respectively. A z-step of 600 nm were used for the z-stack. The scan volume is $50 \times 50 \times 30 \mu m^3$.

### 4.3- Deconvolution of images:

Fluorspheres of 0.170 µm in diameter (Azide 2mM) were introduced into a PMMA solution at a concentration similar to 6T@BNNTs using the same protocol as previously described in section 3. A z-stack fluorescence datacube was performed to recover the PSF of individual fluor sphere. The excitation wavelength was 532 nm and the collection spectral range was 560-631. Image processing included a Z projection Max intensity, a 50 pixel rolling ball to eliminate the background, a 3D Med kernel 1 pixel. The deconvolution Lab 2 [3] plugin on Fiji was used, applying a Richardson Lucy algorithm with 20 iterations on the different datasets. The results of the image processing are shown in Figure S3. Deconvoluted nanotube data (intensity profile) extracted from FIJI software were processed with SciDAVis software. A multi peak Gaussian fitting was performed on each cluster studied, a coefficient of determination $r^2$ between 0.97 and 0.99 was obtained. The non-clustered tubes were also analysed with SciDAVis to calculate the area under the intensity profile.

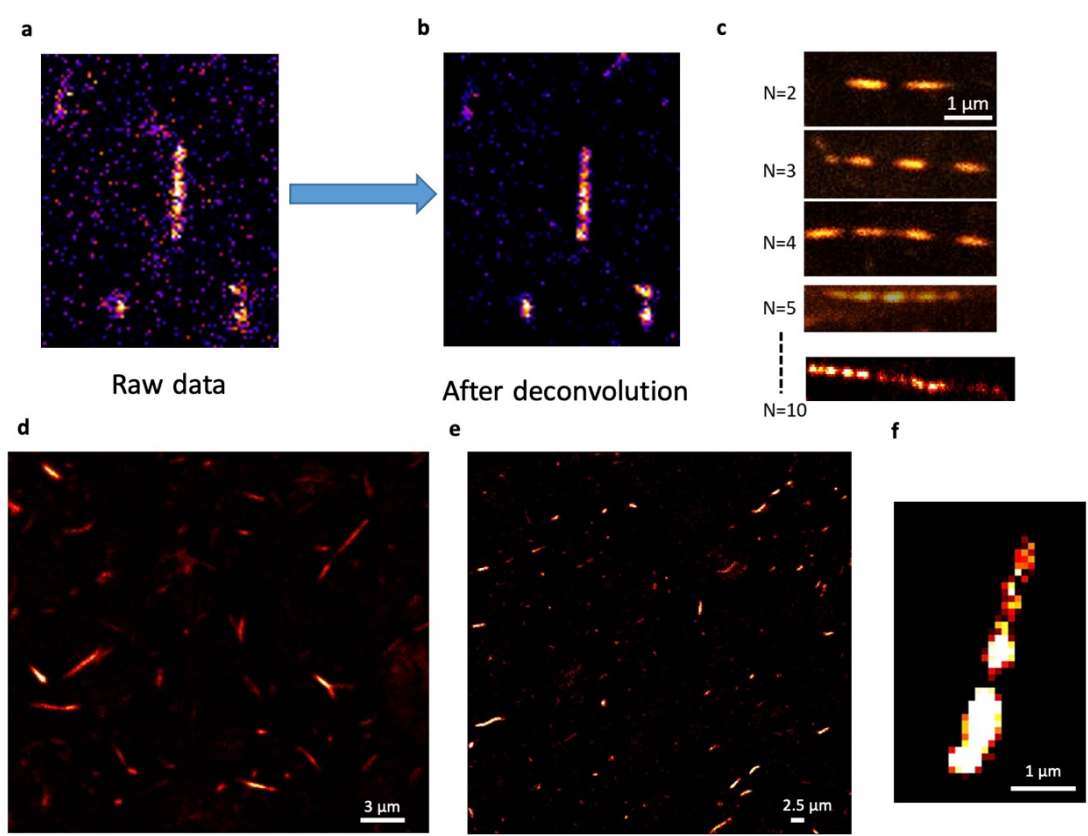

**Figure S3.** (a) Image from stretched PMMA-6T@BNNT by confocal imaging. (b) Same image but after deconvolution step. (c) Zoology of a number of clusters in PMMA-6T@BNNT (d) Confocal image from PMMA-6T@BNNT unstretched, no heating (e) Confocal image from PMMA-6T@BNNT unstretched only heated at 150°C (f) Confocal image from stretched PDMS-6T@BNNT.

## 5. Super-resolved imaging

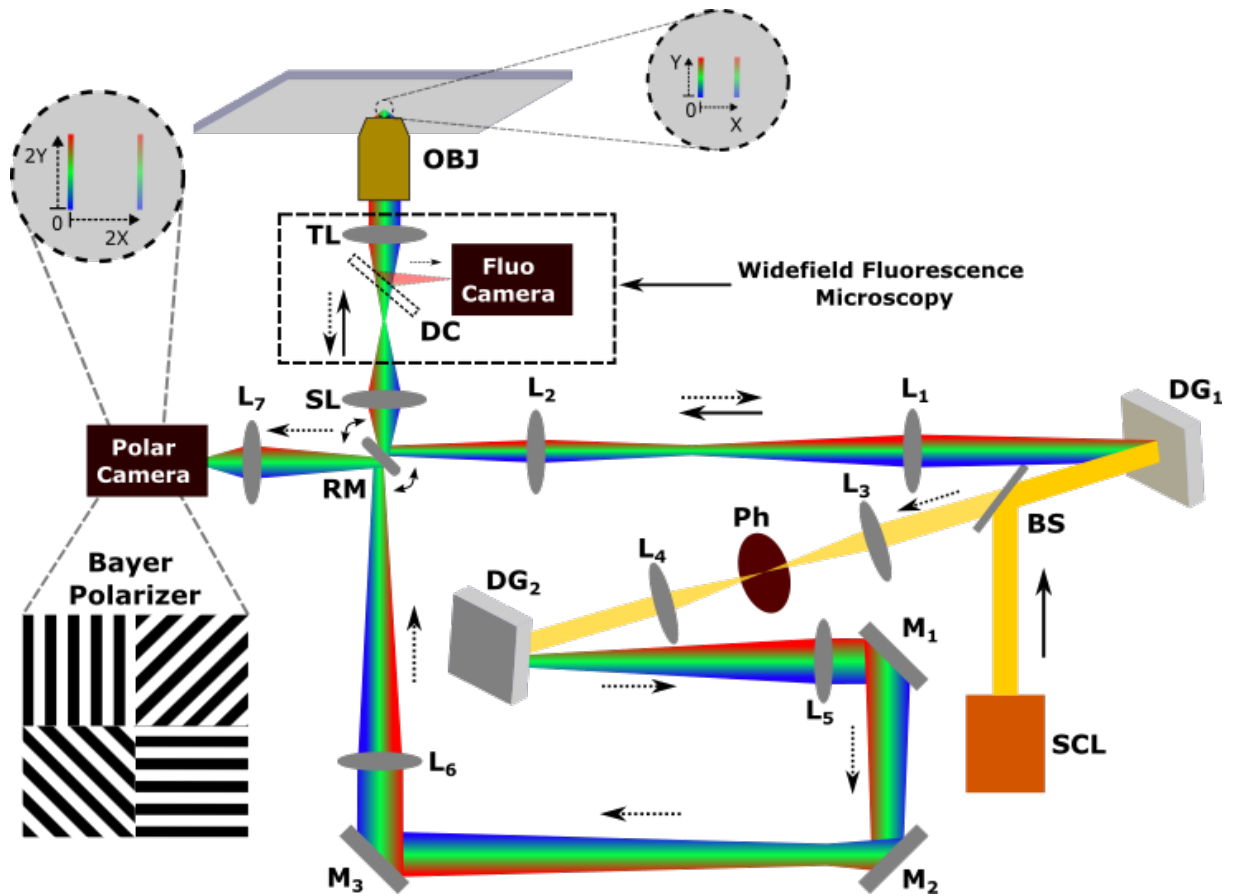

**Figure S4. Super-resolved reflectance system. M: mirror, L: doublet lens, SCL: super-continuum laser, DG : diffraction grating, SL : scanning lens, RM : resonant mirror, OBJ: microscope objective, TL: tube lens, DC: dichroic mirror, Ph: pinhole, BS: beam-splitter.**

The label-free images of **6T@BNNTs** were acquired using super-resolve reflectance system as described in **figure S4** and a previous publication[4]. The system is a confocal-based microscopy where the Image Scanning Microscopy (ISM) technique is implemented to achieve high speed label-free super-resolved imaging. The non-polarized white beam from a super continuum laser is spread into chromatic line by a diffraction grating (DG1). The chromatic line is projected to sample by the objective (OBJ, 60X, water immersion, NA=1.27) and scan by a resonant mirror (RM, @12kHz). The back-scattered signal is collected by the same objective and travel back to the diffraction grating DG1. The pinhole (Ph) is placed in a conjugated image plane to reject the background signal. The reflected signal is spread again by second a diffraction grating (DG2) and rescan by the other face of the resonant mirror. A polarisation sensitive camera (Alkeria, Celera C5S-MP) is used to acquire images of different polarization states at the same time. The wavelength at the centre of image is 550nm where the (super-) resolved PSF is measured at 170 nm (resolution of 120 nm). To acquire fluorescence widefield image for comparison, a dichroic (DC, FF556-SDi01) is inserted between Tube Lens (TL) and Scan Lens (SL). The excitation band is 525-556 nm whereas the collected signal is filtered by long pass filter at 561nm (BLP02-561R). To obtain polarized resolved fluorescence images, a polariser is inserted on the excitation beam before the dichroic align (sequentially) along the 0 and 90° orientation of the tube and one acquisition is performed for each position on a sCMOS camera (Orca Flash 4, Hamamatsu).

The sample was fixed and mounted on a microscope slide in the same way as in confocal fluorescence imaging (section 4.1).

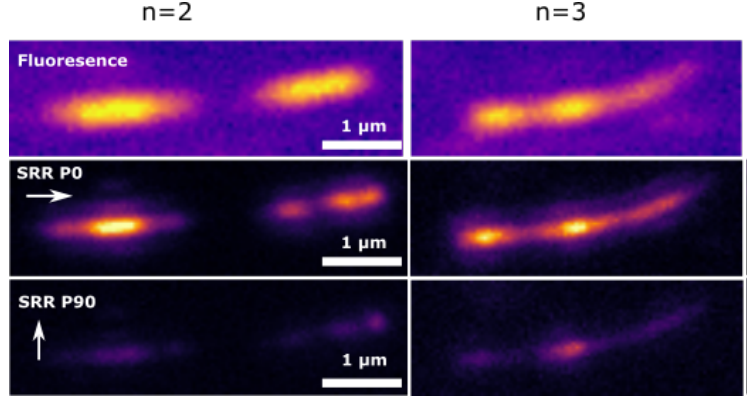

**Figure S5: Fluorescence and Super-resolved reflectance (SRR) polarized images for different tubes exhibiting different cluster numbers.**

6. **Persistence length analysis on BNNT, introduction to static persistence length Lsp**

**6.1- Persistence length analysis on BNNT**

The length of BNNTs is linked to its ability to deform by its stiffness. The stiffness of a tube is directly related to its Young's modulus, its diameter and its length by the relationship[5,6]:

$$l_p = \frac{\pi}{64}\frac{Ed^4}{k_bT} \qquad (S.Eq\ 1)$$

With $l_p$ the persistence length, $k_b$ is Boltzmann's constant and T is the temperature in Kelvin and E the Young's modulus of BNNT and d the inner diameter of BNNT. Therefore a tube of Length (L)< $1.5l_p$ is rigid and does not deform. Tubes equivalent to $l_p$ or higher than $l_p$ will deform.

We calculate a range of $l_p$ based on the experimental data. For diameter in the range of 0.7-8nm. By taking E=1 TPa[7,8] and T=423.15K we obtained $l_p$ =2µm to $l_p$ =34 mm. The larger distribution of diameter in our BNNT batch is at 1.5nm which is corresponding to $l_p$ =44µm. Thermal fluctuation can bend nanotube like a semiflexible polymer if its contour length exceeds the Kuhn statistical segment length of $2l_p$[9]. In our case, 92% of the nanotubes are below $2l_p$. In this case, we can consider 6T@BNNTs as semi-flexible, but only thermal fluctuation cannot be responsible for the bending of 6T@BNNTs in the PMMA matrix at 150°C. 6T@BNNTs length and diameter are key parameters in the understanding of the molecular reorganization.

### 6.2- Static persistence length *lsp*.

Based on the work by Lee et al, [10] the statistical quantity *lsp* can be derived from the spatial average of the square end-to-end vector $\langle R^2 \rangle$ and the bending ratio $D_b$ of a rigid random rod coil formed by randomly distributed static bend points ($\{\phi\} = (\phi_1, \phi_2, ..., \phi_k)$) along the BNNT axis[10], as described in Equations 1a and 1b:

$$\langle R^2 \rangle = N^2 \sum_{i=1}^{k} \sum_{j=1}^{k} (\phi_i \cdot r_i) \cdot (\phi_j \cdot r_j) = N^2 b^2 D_b \quad \text{(S. Eq. 2)}$$

Where $N$ is the total number of unit segments, $k$ is the number of static bending points on a coil, b is and $r_i$ is the i-direction segment vector with the length of $b$.

Assuming constant segment length for renormalization[11] ($\phi_i = \frac{2l_{p0}}{L}$ and $k = \frac{L}{2l_{p0}}$) and small angle ($\theta$) approximation between adjacent segments, $D_b$ can be expressed as Equation 2:

$$D_b \equiv \frac{\langle R^2 \rangle}{L^2} \simeq \left(\frac{2l_{p0}}{L}\right)\left(\frac{1+\cos(\theta)}{1-\cos(\theta)}\right) = M\left(\frac{2l_{p0}}{L}\right) = \frac{2l_{sp}}{L} \quad \text{(S.Eq. 3)}$$

Where $lsp = Ml_{p0}$ is the static bending persistence length, $l_{p0}$ constant segment length and $M$ a constant for a fixed bend angle that depends on the BNNT stiffness.

Hence, we determined the distribution of *lsp* $_{6T@BNNT}$ values in our samples by measuring $\langle R^2 \rangle_{6T@BNNT}$ and $L_{6T@BNNT}$ from a statistical set of 19 curved 6T@BNNTs, as presented in Figure 2e. From this distribution, we defined *lsp* $_{-min}$, *lsp* $_{-max}$ and *lsp* $_{-med}$ as the extrema and median values of $lsp_{-6T@BNNT}$. We also measured the Full Width at Half-Maximum (FWHM) of the cluster's signal as an indicator of their length (Figs 1c and 3a).

In Figure S6a, we can see that periodicity of cluster is correlated to the length of the 6T@BNNT. However, the size of the cluster is not linked with the length of 6T@BNNTs. We formulated the length composition of a clustered 6T@BNNT ($L_{6T@BNNT} = N_{cl} * L_{cl} + N_{gap} * L_{gap}$) as a function of Lsp-6T@BNNT, as described in S.Equation (4):

$$L_{6T@BNNT} = N_{cl} * Lsp_{6T@BNNT} + \varepsilon(N_{cl} - 1) * Lsp_{6T@BNNT} \quad (S.Eq\ 4)$$

$$N_{cl} = \frac{L_{6T@BNNT} + \varepsilon Lsp_{-6T@BNNT}}{Lsp_{-6T@BNNT}(1+\varepsilon)} \quad (S.Eq\ 5)$$

where we assume $N_{gap} = (N_{cl} - 1)$ and where $\varepsilon = 0.3$ is the median value from experimental ratio between $L_{cl}$ and $L_{gap}$, data are available in (c) of figure S6.

We studied the number of cluster per 6T@BNNT as a function of the BNNT length. Following the curvature-induced active diffusion model, we compared, as a function of the BNNT length, the number of cluster observed with the number of cluster predicted by Equation S.Eq 4 using values of lsp$_{min}$, lsp$_{median}$ and lsp$_{max}$. The results are shown in Figure S6d. We observe that these simulations bound well the experimental results. In conclusion, the number of clusters is directly related to the BNNT length and its static persistence length with a numerical prefactor of order 1.

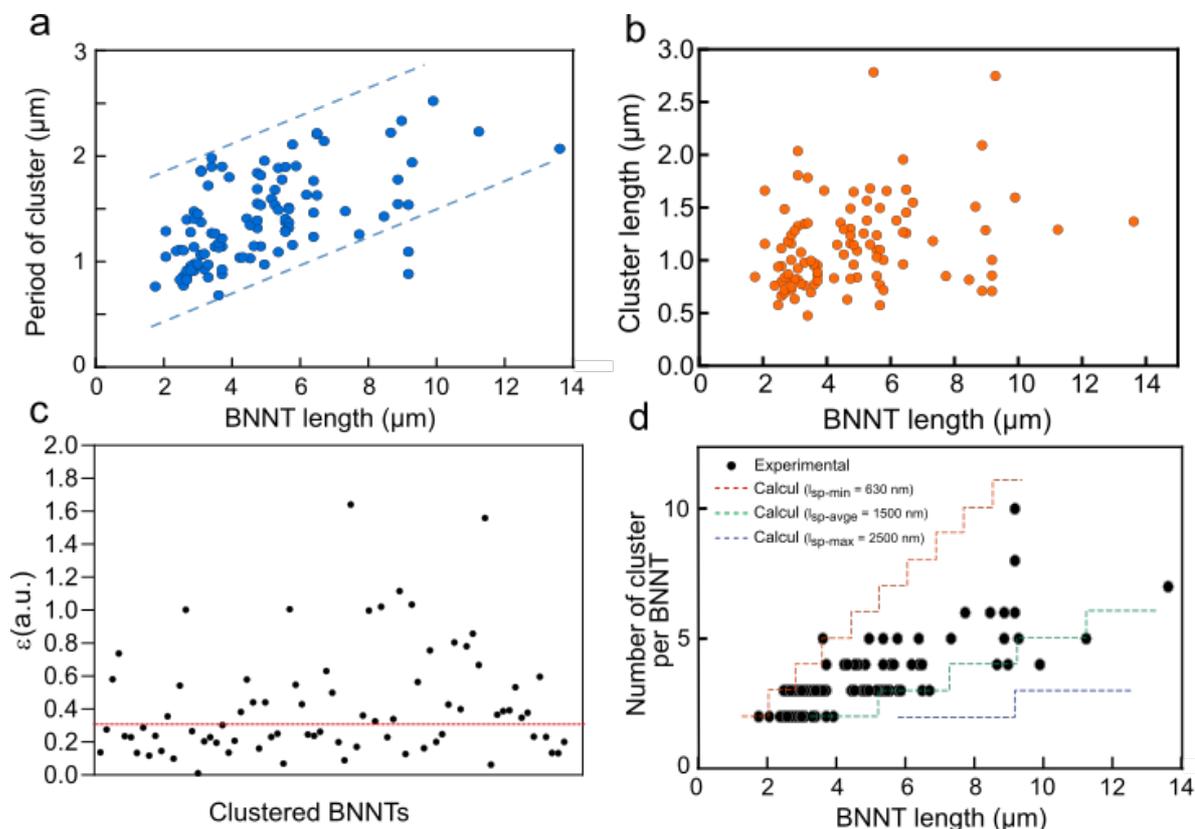

**Figure S6.** (a) Cluster periodicity on each clustered 6T@BNNT as a function of the BNNTs length. (b) Length of the cluster aggregate as a function of the BNNTs length. (c) $\varepsilon$ the ratio between $L_{cl}$ and $L_{gap}$ of the clustered BNNTs (d) Experimental measurement and modelling of the number of clusters per 6T@BNNT as a function of the BNNTs length.

### 6.3 Extraction of the BNNT orientation and local curvature.

The curvature was extracted using the Kappa plugin available in the Fiji software. The parameters used to extract the curve were a B-Spline curve with a point fitting algorithm based on point distance minimization.

The study of the orientation of the tubes with respect to the tensile direction of the matrix was carried out manually using the Fiji software for all the 6T@BNNTs in a volume of PMMA-6T@BNNT film 50x50x2.4µm³ stretched and heated.

### 7. K-mean Analysis of the 6T@BNNT intensities

The average intensity data per cluster in tubes is derived from the intensity profiles extracted from Fiji. Each intensity peak is integrated in a clustered nanotube. We obtain the average intensity/unit length for each cluster. All the peak intensities in a tube are averaged to give the average intensity in a clustered tube, which is physically the representation of the emission intensity of the local molecule density. For the study we have 85 clustered tubes which represents the population of this study.

Using the sklearn.cluster.KMeans library with the parameters: model = KMeans(n_clusters = n, init='k-means++', max_iter=500, random_state=42). (n varying from 1 to 10). By fitting with the average intensities we obtain the inertia value for each groupe n available in figure S7. Using the KElbowVisualizer function in the Yellowbrick library, we can locate the number of groups for which the inertia starts to decrease linearly. The inertia is calculated by measuring the distance between each data point and its centroid, squaring this distance, and summing these squares across one group.

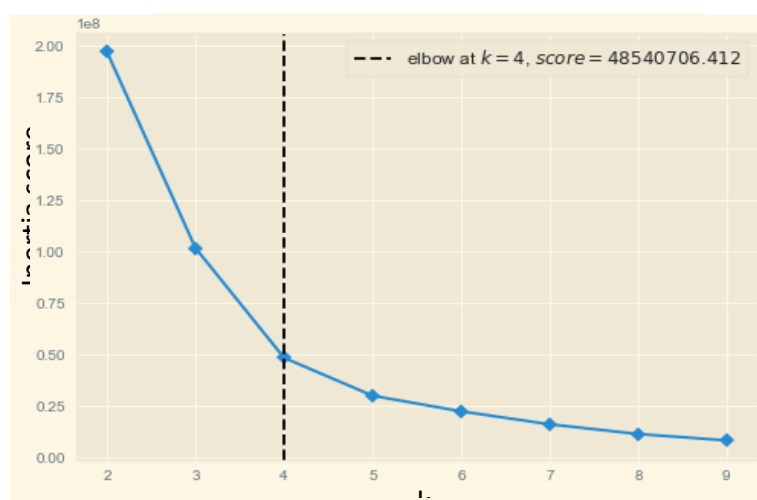

**Figure S7. Inertia curve versus number of groups k**

A good representation of the number of groups can also be shown, using the Silhouette Coefficient. The score is computed by averaging the silhouette coefficient for each sample, computed as the difference between the average intra-cluster distance and the mean nearest-cluster distance for each sample, normalized by the maximum value. This produces a score between 1 and -1, where 1 is highly dense clusters and -1 is completely incorrect clustering. The representation for 2,3,4,5,6 groups is given in figure S8. Note that it is difficult to choose between 4 and 5 groups with this representation. It is assumed that the denser the clusters, the better they are separated, so for a score greater than 0.7 for each of the clusters and in accordance with the elbow analysis, we have divided the clustered nanotubes into four groups. Finally, we plot the distribution of clusters according to the 4 groups determined, available in figure S8.

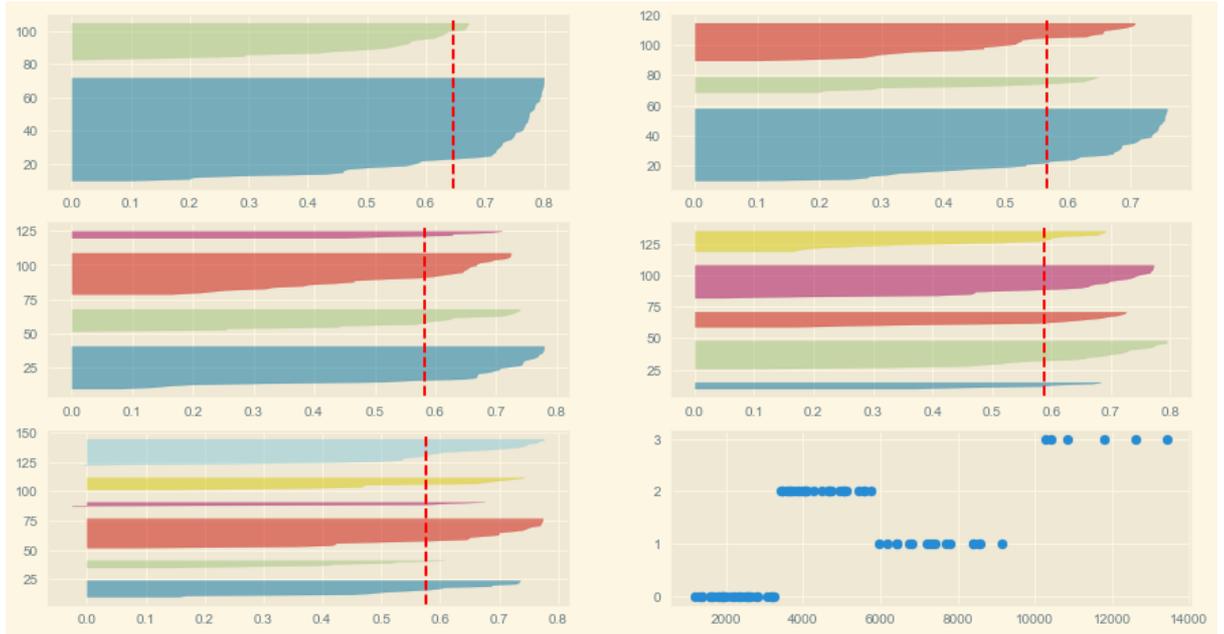

**Figure S8. (a-e) Representation of silhouette coefficient for 2,3,4,5,6 groups. (f) Distribution in average intensity of the 85 clustered BNNT for each of the four groups.**

8. Slope Analysis of the fluorescence intensity profile in 6T@BNNTs (Figure 2)

The aim of this study is to determine the slope coefficient of the tangent at the FWHM abscissa for each intensity profile of a given cluster. This study is based on a statistical assembly of 320 clusters. It was carried out with Python. To do this, we use the formula for the tangent of a differentiable function. If *f* is a function that can be derived on an interval containing a real number a, the tangent to the curve representing *f* at the point with abscissa a has the S.Equation (6):

$$y = f'(a)(x - a) + f(a) \qquad (S.Eq\ 6)$$

with the function *f(x)* being a sum of Gaussian functions that can be derived everywhere and depends on the number of light segments in the clustered nanotube. Thus, for a nanotube with 3 clusters, we have :

```
def f(x):
    return(y0+((2/m.pi))**0.5*A1/w1*np.exp(-2*(x-xc1)**2/w1**2)+((2/m.pi))**0.5*A2/w2*np.exp(-2*(x-xc2)**2/w2**2)+((2/m.pi))**0.5*A3/w3*np.exp(-2*(x-xc3)**2/w3**2))
```

$A_n$ corresponds to the maximum amplitude of the Gaussian, $xc_n$ is the abscissa at the maximum amplitude. Finally $w_n$ is the width. These data are recovered from the Gaussian multipeak fit performed with SciDavis for each tube. We then use a dichotomy algorithm to determine the abscissa at FHWM, denoted x0. Then we calculate the function f at x0. We also calculate the derivative of the function f(x) using the derivate() function from the scipy.misc library. We then obtain the slope of the tangent at the abscissa of FWHM as the derivative f'(a). The set of slopes is given in Figure S9.

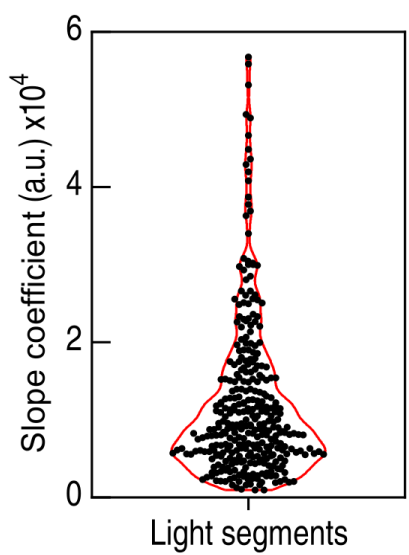

**Figure S9. Representation of the slope coefficient value of the 320 light segment in the 85 clustered 6T@BNNT**

**References:**

1. Tamayo, A. B., Tantiwiwat, M., Walker, B. & Nguyen, T.-Q. Design, Synthesis, and Self-assembly of Oligothiophene Derivatives with a Diketopyrrolopyrrole Core. *J. Phys. Chem. C* **112**, 15543–15552 (2008).
2. Allard, C. *et al.* Confinement of Dyes inside Boron Nitride Nanotubes: Photostable and Shifted Fluorescence down to the Near Infrared. *Advanced Materials* **32**, (2020).
3. Sage, D. *et al.* DeconvolutionLab2: An open-source software for deconvolution microscopy. *Methods* **115**, 28–41 (2017).
4. Ta, D.-M., Aguilar, A. & Bon, P. Label-free image scanning microscopy for kHz super-resolution imaging and single particle tracking. *Opt. Express, OE* **31**, 36420–36428 (2023).
5. Flexural rigidity of microtubules and actin filaments measured from thermal fluctuations in shape. *J Cell Biol* **120**, 923–934 (1993).
6. Landau, L.D. & Lifshitz, E.M. *Statistical Physics*. vol. p. §127 (Pergamon Press, 1969).
7. Arenal, R., Wang, M.-S., Xu, Z., Loiseau, A. & Golberg, D. Young modulus, mechanical and electrical properties of isolated individual and bundled single-walled boron nitride nanotubes. *Nanotechnology* **22**, 265704 (2011).
8. Kim, J. H., Pham, T. V., Hwang, J. H., Kim, C. S. & Kim, M. J. Boron nitride nanotubes: synthesis and applications. *Nano Convergence* **5**, 17 (2018).
9. Sano, M., Kamino, A., Okamura, J. & Shinkai, S. Ring Closure of Carbon Nanotubes. *Science* **293**, 1299–1301 (2001).
10. Lee, H. S., Yun, C. H., Kim, H. M. & Lee, C. J. Persistence Length of Multiwalled Carbon Nanotubes with Static Bending. *J. Phys. Chem. C* **111**, 18882–18887 (2007).
11. Gennes, P.-G. de. *Scaling Concepts in Polymer Physics*. (Cornell University Press, 1979).